\documentstyle[psfig,frontieres,art10]{article} 
\def\be{\begin{equation}}
\def\ee{\end{equation}}
\def\bea{\begin{eqnarray}}
\def\eea{\end{eqnarray}}

\begin{document} 
\heading{%
%
Removal of interference from
gravitational wave spectrum
%
}  
\par\medskip\noindent 
\author{
Alicia M. Sintes$^{1}$, Bernard F. Schutz$^{1,2}$ 
} 
\address{
Max-Planck-Institut f\"ur Gravitationsphysik (Albert-Einstein-Institut),
Schlaatzweg 1, D-14473 Potsdam, Germany
} 
\address{
Department of Physics and Astronomy, University of Wales College of Cardiff,
U.K.
} 
%
 
\begin{abstract} 
We develop a procedure 
to remove interference from gravitational wave spectrum.
 The method is applied to the data 
produced by the Glasgow laser interferometer in 1996 and all the
lines corresponding to the interference with 
the main supply are removed.
\end{abstract} 
\section{Introduction} 
In this paper we present  an algorithm  to remove  interference from the
gravitational wave ({\sc gw}) spectrum. 
This method allows the removal  of coherent lines
coming from deterministic signals while keeping the intrinsic detector
noise. 
Unlike other existing methods for removing single interference lines
 \cite{A,Th}, the method described here can remove the external  interference 
 without removing any gravitational wave signal that  may be hidden by the
 interference.
Therefore, it can
be very useful in the search for monochromatic {\sc gw} 
signals as those ones produced by pulsars
 \cite{S,T}.

The key to this method is to determine the interference  by using many 
harmonics of the interference signal. In the study of the data produced 
 by the Glasgow laser interferometer in March 1996
\cite{J},
we observe in the spectrum many instrumental lines, some of them at
multiples of 50 Hz. All these lines are wide, and when we compare them,
we observe that their overall structure  is very similar 
but only  the scaling of the width is different (see figure 1).
\begin{figure} 
\centerline{\vbox{ 
\psfig{figure=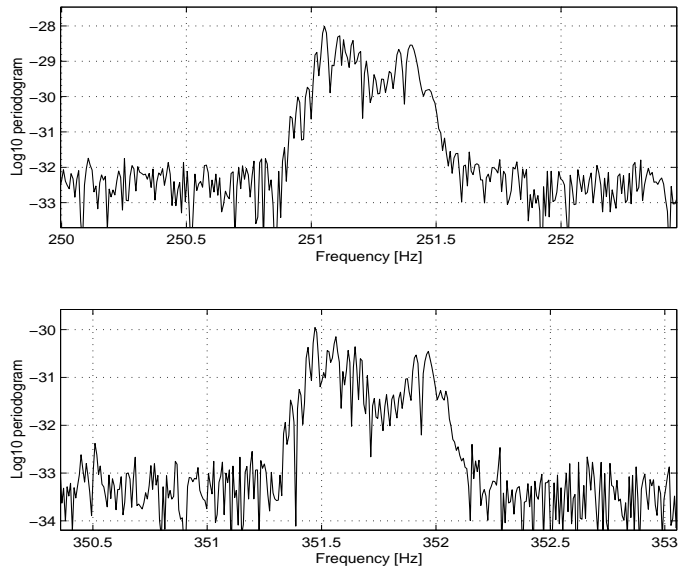,height=7.5cm,width=9.0cm} 
}} 
\caption[]{Comparison of the structure of the lines at 250 Hz and at 350 Hz
of the power
spectrum of the Glasgow data.  
The broad shape is due to the wandering of the incoming electricity 
frequency.
} 
\end{figure} 
If we look at these lines in more detail, in smaller length Fourier transform
(seconds in length), they appear as well defined
small bandwidth lines which change frequency over time
in the same way, while other ones
 remain at constant frequency. Therefore, all these lines at multiples
of 50 Hz must be harmonics of a single source, for example the main supply.

 In the Glasgow data, the lines at 1 KHz have a width of
5 Hz. Therefore, we can ignore these sections of the power spectrum or
we can try to remove this interference in order to be able to detect
{\sc gw} signals hidden behind those lines.
The LIGO group has also reported largely instrumental artifacts
 at multiples of the 60 Hz line frequency in their  
 40 m interferometer \cite{Abra}. Therefore, this seems to be a general effect
 present in the different interferometers.
 
 This electrical interference may possibly be reduced by 
 improvements in the  electronic
design that will be incorporated  in the next generation  of detectors.
However, we cannot be sure that such interference will be completely absent
or that other
sources of interference will not manifest themselves in long-duration
spectra. Indeed, the Glasgow data \cite{J} contain other regular
features of unknown origin \cite{K}. 
For this reason we
 investigate  solutions to the problem  using the existing data.

We address here only the removal of coherent lines, not stochastic ones
(such as those due to thermal noise). Our method requires 
coherence between the fundamental and several harmonics. If there is no
such coherence, other methods \cite{A,Th} can be used, but these will
remove {\sc gw} signals too. The method presented here is the only one
we know that allows one to find real signals under the interference.

\section{The basic concept} 

We suppose that a electrical signal at $\sim$~50 Hz enters in the 
electronics of the interferometer. The instrument then 
transforms the input signal into a series of harmonics in a
stationary way. Therefore, the data stream contains the function
$y(t)$ given by
\be
y(t)=\sum_n a_n m(t)^n + \left( a_n m(t)^n\right)^* \ ,
\label{e3}
\ee
where $a_n$ are constants and $m(t)$ is  nearly monochromatic around
50 Hz. 
The total output, $x(t)$, contains also  the gravitational wave 
signals $s(t)$ and the detector noise $n(t)$
\be
x(t)=y(t)+n(t)+s(t)\ .
\label{e1}
\ee

We know that the data recorded is band-limited since 
an anti-aliasing filter was applied to the data before it was sampled.
 Therefore, the function $y(t)$ must be also band-limited,
and hence, the sum can be limited to the first 39 harmonics. This number
is given by the Nyquist frequency that is related to the sampling 
frequency of the experiment ($
n_{max}=f_{Nyquist}/50 \ {\rm Hz}- 1$).

Our purpose is to construct a function
\be
h(t)=\sum_{n=1}^{39} \rho_n M(t)^n + \left( \rho_n M(t)^n\right)^* \ ,
\label{e5}
\ee
similar to $y(t)$. Thus, we have to determine the complex function $M(t)$
and all the parameters $\rho_n$. Notice that from the experimental data 
we do   not independently know the value of the input signal $m(t)$.

\section{The algorithm} 
In this section we present the algorithm to remove  interference
from the gravitational wave spectrum. 

As we pointed out in the previous section, we assume that the data produced
by 
the interferometer is just the sum of the interference  plus the detector
noise and possible {\sc gw} signals that we will not consider  here
\be
x(t)=y(t)+n(t) \ ,
\label{e6}
\ee
where $y(t)$ is given by Eq.~(\ref{e3}). Therefore, the Fourier transform
of the data $\tilde x(\nu)$ is simply given by
\be
\tilde x(\nu)=\tilde y(\nu)+\tilde n(\nu) \ .
\label{e7}
\ee

First, we select a set of harmonics 
(in our case the odd harmonics were strong, i.e., $k=[3,5,7,9,\ldots]$),
or all of them if we prefer. The idea is to construct the function
$M(t)$ getting the maximum information inbeded in the harmonics considered.
For each of these harmonics, we determine the initial and final
frequency of the line $(\nu_{ik}, \nu_{fk})$  and we define a set of functions
$\tilde z_k(\nu)$ in the frequency domain as
\be
\tilde z_k(\nu)= \left\{
\begin{array}{cc}
\tilde x(\nu) & \nu_{ik}<\nu <\nu_{fk}\\
0 & \mbox{elsewhere}\ .
\end{array}
\right.
\label{e8}
\ee
Comparing Eq.~(\ref{e8}) with (\ref{e3}) and (\ref{e7}) we have
\be
\tilde z_k(\nu)= a_k \widetilde{m^k} +\tilde n_k(\nu) \ ,
\ee
where
\be
\tilde n_k(\nu)= \left\{
\begin{array}{cc}
\tilde n(\nu) & \nu_{ik}<\nu <\nu_{fk}\\
0 & \mbox{elsewhere}\ ,
\end{array}
\right.
\ee
is considered to be a zero-mean random complex noise.

Then, we calculate  their inverse Fourier transform
\be
z_k(t)=a_k m(t)^k +n_k(t) \ ,
\ee
and we define 
\be
B_k(t)\equiv \left[ z_k(t)\right]^{1/k}= (a_k)^{1/k}m(t) \beta_k(t) \ ,
\ee
where
\be
\beta_k(t)=\left[ 1+ {n_k(t) \over a_k m(t)^k}\right]^{1/k} \ .
\ee
All these function $\{B_k(t)\}$ are almost monochromatic at $\sim\,$50 Hz,
but they have different  amplitude and phase. In order to compare them,
 we construct another set of functions
 \be
 b_k(t)=\alpha m(t) \beta_k(t) \ ,
 \ee
by performing  the operation
\be
 b_k(t)=\Gamma_k B_k(t)\ , \qquad 
 \Gamma_k=
 \sum_j B_1(j\Delta t) B_k(j\Delta t)^*{\Big
 { /}}
 \sum_j \vert B_k(j\Delta t)\vert^2 \ .
\ee
These new functions $\{b_k(t)\}$ form a set of random variables $-$functions 
of
time$-$ and  they all have the same mean value
\be
\langle b_i\rangle=\alpha m(t) \ .
\ee

\begin{figure} 
\centerline{\vbox{ 
\psfig{figure=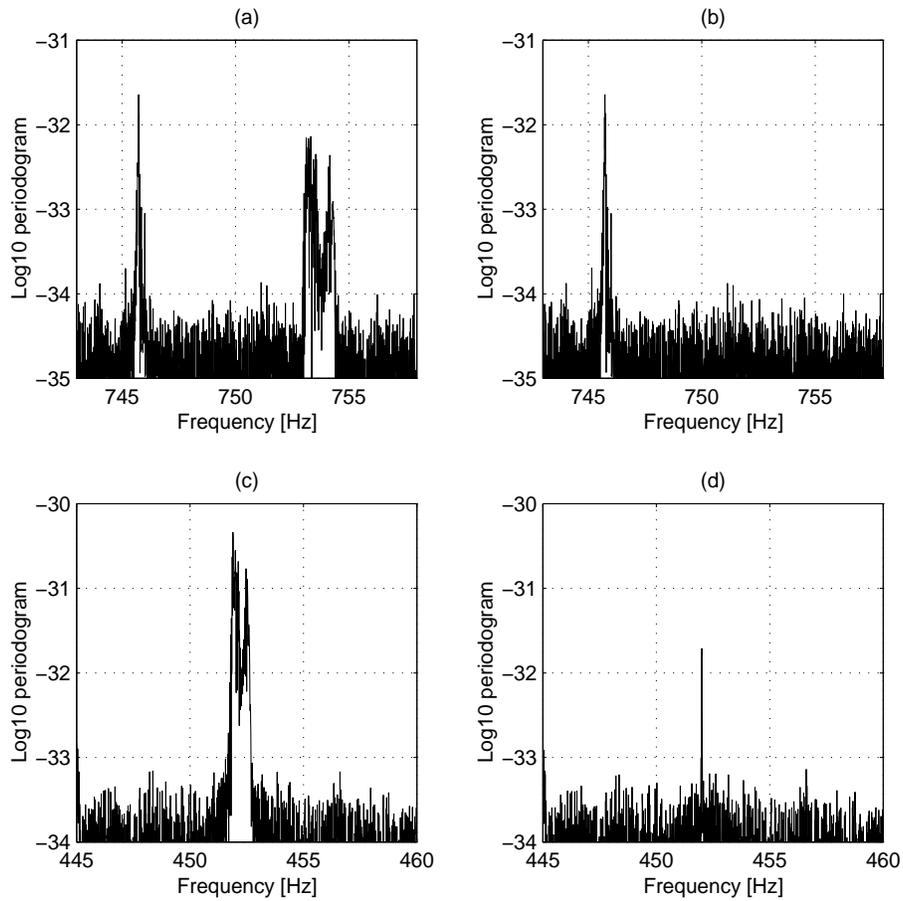,height=12.0cm,width=12.0cm} 
}} 
\caption[]{Decimal logarithm of the
periodogram of 128 blocks (approximately 2 minutes) of the Glasgow data. 
(a) One of the harmonics near 754 Hz.
(b) The same data after the removal of the interference as described 
in the text.
(c)  The same experimental  data
with an artificial  signal added at 452 Hz. 
(d) The data in (c) after the removal of the interference, revealing that
the signal remains detectable. Its amplitude is hardly changed by removing
the interference. 
} 
\end{figure} 

Now, we want to construct $M(t)$ as a function of all $\{b_k(t)\}$,
 in such a way
that it has the same mean and minimum variance. If we assume the function
$M(t)$ to be linear with $\{b_k(t)\}$, the statistically the best is
\be
 M(t)=\left(\sum_k {b_k(t) \over {\rm Var}[\beta_k(t)]} \right) {\Big
 { /}}
\left( \sum_k {1 \over {\rm Var}[\beta_k(t)]}\right) \ ,
\ee
where
\be
{\rm Var}[\beta_k(t)]= {1\over k^2}{\langle n_k(t) n_k(t)^*\rangle\over 
\vert a_k m(t)^k\vert^2}+ \mbox{corrections} \ ,
\ee
\be
a_k m(t)^k\approx z_k(t) \ ,
\ee
\be
\langle n_k(t) n_k(t)^*\rangle=\int d\nu \int d\nu'
\langle \tilde n_k(\nu) \tilde n_k(\nu')^*\rangle e^{2 \pi i(\nu-\nu')t} \ .
\ee
If we assume the noise to be stationary  (i.e., 
$\langle \tilde n(\nu) \tilde n(\nu')^*\rangle= S(\nu) \delta(\nu-\nu')$),
the previous equation becomes
\be
\langle n_k(t) n_k(t)^*\rangle = 
\int d\nu \int d\nu' \, S_k(\nu) \delta(\nu-\nu') e^{2 \pi i(\nu-\nu')t}
=\int_{\nu_{ik}}^{\nu_{fk}} S(\nu) d\nu \ ,
\ee
where $S(\nu)$ is the power spectral density of the noise.

Finally, it only remains to determine the parameters $\rho_n$ in Eq.~(\ref{e5})
that we obtain by performing the operation
\be
\rho_n=
 \sum_j x(j\Delta t) M^n(j\Delta t)^*{\Big
 { /}}
 \sum_j \vert M^n(j\Delta t)\vert^2 \ .
\ee

We have applied this method to the data taken from the Glasgow
laser interferometer in 1996 and we have succeeded in removing the electrical
 interference. The same method has
also been applied  to the true
experimental data with an external simulated signal at 452 Hz, that
remains hidden due to its weakness, and we have succeeded in removing the
electrical interference while keeping the signal present in the data,
obtaining a clear outstanding peak over the noise level
(see figure 2). This will be described in detail elsewhere \cite{ali}.

This method can also be applied to any other kind of interference and
it may have more applications, not only for the detection of 
 {\sc gw} radiation, but also, for example, in radioastronomy.

\acknowledgements{This work was supported by the European Union, 
 TMR Contract
No. ERBFMBICT961479.}

\vfill 

\begin{iapbib}{99} 
\bibitem{Abra} Abramovici A., Althouse W., Camp J., Durance D.,
Giaime J.A., Gillespie A., Kawamura S., Kuhnert A.,
Lyons T., Raab F.J., Savage Jr. R.L., Shoemaker D., Sievers L.,
Spero R., Vogt R., Weiss R., Whitcomb S., Zucker M.,
1996, Physics Letters A 218, 157
\bibitem{A} Allen B., {\em Data analysis package GRASP},
in this volume
\bibitem{J} Jones G.S., 1996, {\em Fourier analysis of the data produced
by the Glasgow laser interferometer in March 1996}, internal report,
Cardiff University, Dept. of Physics and Astronomy
\bibitem{K} Krolak A., {\em Statistical test for periodical signals}, 
in this volume
\bibitem{S} Schutz B.F., 1997, eds. Marck J.A. \& Lasota J.P., in 
{\it Relativistic Gravitation and Gravitational Radiation}. Cambridge
University Press, Cambridge
\bibitem{ali} Sintes A.M., Schutz B.F., to be published
\bibitem{Th} Thomson D.J., 1982, {\em Spectrum Estimation and 
Harmonic Analysis} in {\it Proceedings of the IEEE}, 70, 1055-96
\bibitem{T} Thorne K.S., 1995, eds. Kolb E.W. \& Peccei R., in
{\it Proceedings of Snowmass 1994 Summer Study on Particle and Nuclear 
Astrophysics and Cosmology}. World Scientific, Singapore, p. 398
\end{iapbib}
\end{document}